\documentclass{PoS}

\newcommand{\kslash}{k\kern-1ex /}
\newcommand{\pslash}{p\kern-1ex /}
\newcommand{\qslash}{q\kern-1ex /}
\newcommand{\lslash}{l\kern-1ex /}
\newcommand{\sslash}{s\kern-1ex /}
\newcommand{\Dslash}{D\kern-1.2ex /}

\newcommand{\beqa}{\begin{eqnarray}}
\newcommand{\eeqa}{\end{eqnarray}}

\newcommand{\bd}{\begin{description}}
\newcommand{\ed}{\end{description}}

\newcommand{\ben}{\begin{eqnarray}}
\newcommand{\een}{\end{eqnarray}}

\def\lsim{\raise0.3ex\hbox{$<$\kern-0.75em\raise-1.1ex\hbox{$\sim$}}}
\def\gsim{\raise0.3ex\hbox{$>$\kern-0.75em\raise-1.1ex\hbox{$\sim$}}}
\def\simgt{\rlap{\lower 3.0 pt\hbox{$\mathchar \sim$}}\raise 2.0pt \hbox {$>$}}
\def\simlt{\rlap{\lower 3.0 pt\hbox{$\mathchar \sim$}}\raise 2.0pt \hbox {$<$}}

\newcommand{\msbar}{{\overline {\rm MS}}}

\title{PACS-CS results for 2+1 flavor lattice QCD simulation on and off the physical point}

\ShortTitle{PACS-CS results for 2+1 flavor lattice QCD simulation on and off the physical point}

\author{Yoshinobu Kuramashi\thanks{E-mail: 
kuramasi@het.ph.tsukuba.ac.jp}\hspace{3mm}for the PACS-CS Collabolation \\
Center for Computational Sciences and
Graduate School of Pure and Applied Sciences,\\ 
University of Tsukuba, Tsukuba, Ibaraki 305-8571, Japan}

\abstract{We report on the PACS-CS project focusing on a
direct simulation of 2+1 flavor QCD on the physical point
and chiral analysis of meson and
baryon masses off the physical point with both the
SU(2) and SU(3) chiral perturbation theories.
Configurations are generated with the $O(a)$-improved Wilson quark
action and the Iwasaki gauge
action. The up-down quark is simulated by employing the DDHMC algorithm
with several improvements and the UV-filtered PHMC algorithm
is implemented for the strange quark.
We investigate the convergence behaviors of the SU(2)
and SU(3) chiral expansions up to NLO for 
the pseudoscalar meson sector,
where the up-down quark mass ranges from
3 MeV to 24 MeV  and the strange quark mass is chosen 
around the physical value.
The fit results for the low energy constants are compared with those
recently obtained by other groups.
We also discuss the importance of the direct simulation at the physical point
by comparing the physical quantities measured
on the physical point with those estimated
by the extrapolation method.}

\FullConference{The XXVI International Symposium on Lattice Field Theory\\
		 July 14-19 2008\\
		 Williamsburg, Virginia, USA}

\begin{document}

\section{Introduction}

The aim of lattice QCD calculation is to nonperturbatively evaluate 
physical quantities from first principles with the systematic errors 
under control. Most troublesome are quenching effects 
and uncertainties associated with chiral extrapolation, which should be
removed by 2+1 flavor simulation at the physical point.

The previous CP-PACS/JLQCD project\cite{cppacs_nf3} focuses on 
performing $N_f=2+1$ lattice QCD simulations incorporating the dynamical
strange quark by the Polynomial Hybrid Monte Carlo (PHMC) 
algorithm\cite{phmc}, where the nonperturbatively $O(a)$-improved 
Wilson quark action\cite{csw_np} and the Iwasaki gauge action\cite{iwasaki} are
employed on a (2 fm$)^3$ lattice with the up-down quark mass down to
67 MeV.     
This project is succeeded by the PACS-CS (Parallel Array Computer 
System for Computational Sciences)
project\cite{ukawa1,ukawa2,boku,ishikawa_lat06,kura_lat07,
ukita_lat07,kadoh_lat07,pacscs_nf3,ukita_lat08,kadoh_lat08}, which
aims at the physical point simulation with the same quark and gauge
actions as the CP-PACS/JLQCD project on enlarged physical volumes.
We reduce the up-down quark masses  
using the domain-decomposed HMC (DDHMC) algorithm armored 
with several improvements
and implement the UV-filtered PHMC algorithm\cite{ishikawa_lat06} for 
the strange quark. 

In this report we demonstrate the
feasibility of a direct simulation at the physical point and 
examine viability of extrapolation method with the chiral
perturbation theories.
We apply the SU(2) and SU(3) chiral perturbation theories (ChPTs)
to the pseudoscalar meson masses and the decay constants 
with the up-down quark mass ranging from
3 MeV to 24 MeV and two choices of the strange quark mass
around the physical value.
The low energy constants in the SU(2) and SU(3) ChPTs are determined
and compared with those obtained with the different quark and gauge actions
employed by other groups. 
We discuss the convergence behaviors of the SU(2)
and SU(3) chiral expansions up to NLO.
We also investigate the quark mass dependence of the nucleon mass
employing the SU(2) heavy baryon chiral perturbation theory up to NNLO.
We finally make a comparison between  
the physical quantities directly measured
on the physical point and those estimated
by the extrapolation method. This comparison reveals the necessity and
the effectiveness of the direct simulation at the physical point. 
 
All the calculations are done using the PACS-CS computer
with a total peak speed of 14.3 TFLOPS developed and installed at
University of Tsukuaba on 1 July 2006. A part of the results are already 
reported in Ref.~\cite{pacscs_nf3,ukita_lat08,kadoh_lat08}.

\section{Simulation details}

\begin{table}[h!]
\setlength{\tabcolsep}{10pt}
\renewcommand{\arraystretch}{1.2}
\centering
\caption{Simulation parameters. MD time is the number of
trajectories multiplied by the trajectory length $\tau$.}
\label{tab:param}
\begin{tabular}{llccccc} \hline
$\kappa_{\rm ud}$ & $\kappa_{\rm s}$ & DDHMC  & $\tau$& $(N_0,N_1,N_2,N_3,N_4)$ & MD time & $m_\pi L$ \\ \hline \hline
0.13700  & 0.13640 & plain &0.50& (4,4,10) & 2000 & 10.3 \\
0.13727  & 0.13640 & plain &0.50& (4,4,14) & 2000 &  8.4 \\
0.13754  & 0.13640 & plain &0.50& (4,4,20) & 2250 &  6.0   \\
         & 0.13660 & plain &0.50& (4,4,28) & 2000 &   5.7   \\
0.13770  & 0.13640 & plain &0.25& (4,4,16) & 2000& 4.3 \\
0.13781  & 0.13640 & MP    &0.25& (4,4,4/6,6) & 990 & 2.3   \\ 
0.137785 & 0.13660 & MP2   &0.25& (4,4,2,4,4) & 1000 & 2.4   \\ \hline
\end{tabular}
\end{table}

Our simulation is carried out using the $O(a)$-improved Wilson quark
action with a nonperturbative improvement coefficient $c_{\rm
SW}=1.715$\cite{csw_np} on a $32^3\times 64$ lattice. 
The lattice spacing is found to be 0.0907(13) fm whose determination is
explained later. Simulation
parameters are summarized in Table~\ref{tab:param}. We choose seven
combinations of the hopping parameters $(\kappa_{\rm ud},\kappa_{\rm
s})$ based on the analysis of the previous CP-PACS/JLQCD results.
We employ two choices of $\kappa_{\rm s}$ to investigate the 
strange quark mass dependences.  
The heaviest combination 
$(\kappa_{\rm ud},\kappa_{\rm s})=(0.13700,0.13640)$ corresponds
to the lightest one in the previous CP-PACS/JLQCD project.  
We expect that the combination 
$(\kappa_{\rm ud},\kappa_{\rm s})=(0.137785,0.13660)$ 
would be the physical point, which is estimated at the early stage of
our analyses.

The DDHMC algorithm\cite{luscher} allows us to access the small up-down 
quark mass region closer to the physical point.  It is 
employed for the simulation points with $\kappa_{\rm ud}\le 0.13770$.
We divide the full lattice into $8^4$ blocks which are used for 
a geometric separation of the up-down quark determinant into the UV and
the IR parts.
This separation makes possible to incorporate the multiple time
scale integration scheme\cite{sexton}, which reduce the stimulation cost 
significantly. We employ the replay trick\cite{luscher,kennedy} 
choosing the threshold $\Delta H>2$. 
The detailed description about the DDHMC algorithm is
given in Refs.~\cite{pacscs_nf3,ukita_lat08}.  
 
Reducing the up-down quark mass, however, 
the IR force $F_{\rm IR}$ becomes less stable yielding spike-like fluctuations,
which results in larger $\Delta H$ with higher replay rates. 
At the simulation point of $(\kappa_{\rm ud},\kappa_{\rm s})=(0.13781,013640)$ 
we incorporate  the mass preconditioning\cite{massprec1,massprec2} 
to tame the fluctuations of 
the IR force, which is divided 
into the preconditioner $F_{\rm IR}^{\prime}$ with a new hopping parameter
$\kappa^{\prime}_{\rm ud}$ 
and the preconditioned part $\tilde{F}_{\rm IR}$.
$\kappa^{\prime}_{\rm ud}$ is parametrized as $\kappa^{\prime}_{\rm
ud}=\rho_1\kappa_{\rm ud}$ with $\rho_1$ less than unity.
The step sizes for $F_{\rm G}$, $F_{\rm UV}$, $F_{\rm IR}^\prime$, 
$\tilde{F}_{\rm IR}$ are controlled by four integers
$(N_0,N_1,N_2,N_3)$ as $\delta\tau_{\rm G}=\tau/(N_0 N_1 N_2 N_3)$,
$\delta\tau_{\rm UV}=\tau/(N_1 N_2 N_3)$, $\delta\tau_{\rm IR}^\prime=\tau/(N_2
N_3)$, $\delta{\tilde \tau}_{\rm IR}^\prime=\tau/N_3$ with $\tau$ the
trajectory length. 
This algorithm is refereed to as MPDDHMC.
We choose $\delta \tau_{\rm s}=\delta\tau_{\rm
IR}^\prime$ for the strange quark force in the UVPHMC algorithm
based on our observation that 
$||F_{\rm s}||\approx ||F_{\rm IR}^\prime||$. 

At $(\kappa_{\rm ud},\kappa_{\rm s})=(0.13781,013640)$ we also implement
several improvements for the inversion of the Wilson-Dirac operator
on the full lattice.
First one is the chronological guess for the initial
solutions\cite{chronological} 
with the use of the last 16 solutions. Second one is a nested
BiCGStab solver consisting of the outer solver with double
precision arithmetic and the inner one operated with single precision.
The latter with an automatic stopping condition 
from $10^{-3}$ to $10^{-6}$ works as a preconditioner for the former.
We employ a stringent tolerance $|Dx-b|/|b|< 10^{-14}$ for the outer solver
to retain the reversibility of the molecular dynamics trajectories
to high precision. Third one is the GCRO-DR (Generalized Conjugate
Residual with implicit inner Orthogonalization and Deflated Restarting)
algorithm\cite{deflation} which is robust against the small eigen
values of the Wilson-Dirac operator. 
It take over the inversion once the nested BiCGStab solver 
becomes stagnant. 

For the run at $(\kappa_{\rm ud},\kappa_{\rm s})=(0.137785,013660)$
we apply twofold mass preconditioning (MP2DDHMC) to the IR force 
$F_{\rm IR}$,
which is decomposed into $F_{\rm IR}^{\prime\prime}$, ${\tilde
F}^\prime_{\rm IR}$ and ${\tilde F}_{\rm IR}$.
Two hopping parameters are additionally introduced:
$\kappa_{\rm ud}^{\prime\prime}=\rho_2\kappa_{\rm ud}^\prime=
\rho_2\rho_1\kappa_{\rm ud}$  with $\rho_1$ and $\rho_2$ less than
unity. We need five integers $(N_0,N_1,N_2,N_3,N_4)$ to adjust the 
step sizes for $F_{\rm G}$, $F_{\rm UV}$, $F_{\rm IR}^{\prime\prime}$, 
${\tilde F}_{\rm IR}^{\prime}$, $\tilde{F}_{\rm IR}$. 
Our choice is $\delta \tau_{\rm s}=\delta\tau_{\rm IR}^{\prime\prime}$
for the UVPHMC algorithm.

\section{ChPT analyses on the pseudoscalar meson sector}

We examine the chiral behaviors of the pseudoscalar meson masses 
and the decay constants based on the SU(3) and SU(2) ChPTs up to NLO.
Since the redefinition of some of the low energy constants (LECs)
makes the one-loop expressions of the Wilson ChPT 
in terms  of the AWI quark masses\cite{wchpt} equivalent to those in the
continuum\cite{pacscs_nf3}, we focus on the analyses with the use of 
the continuum ChPTs. 

\subsection{SU(3) ChPT}

The one-loop expressions in the continuum SU(3) ChPT are 
given by\cite{chpt_nf3}
\ben
\frac{m_{\pi}^2}{2 m_{\rm ud}}&=& B_0 \left\{
1+\mu_\pi-\frac{1}{3}\mu_\eta 
+\frac{2B_0}{f_0^2} \left(
16 m_{\rm ud}  (2L_{8}-L_5) 
+16 (2 m_{\rm ud} +m_{\rm s}) (2L_{6}-L_4)  
\right) \right\}, 
\label{eq:chpt_mpi}
\\
\frac{m_K^2}{(m_{\rm ud} +m_{\rm s})}&=&B_0 \left\{
1+\frac{2}{3}\mu_\eta
+\frac{2B_0}{f_0^2}\left(
8(m_{\rm ud}+m_{\rm s}) (2L_{8}-L_5)
+16(2 m_{\rm ud}  +m_{\rm s}) (2L_{6}-L_4) 
 \right)\right\}, 
\label{eq:chpt_mk}
\\
f_\pi &=&f_0\left\{
1-2\mu_\pi-\mu_K
+ \frac{2B_0}{f_0^2} \left (
8 m_{\rm ud} L_5+8(2 m_{\rm ud} +m_{\rm s})L_4
\right)\right\}, 
\label{eq:chpt_fpi}
\\
f_K &=&f_0\left\{
1-\frac{3}{4}\mu_\pi-\frac{3}{2}\mu_K-\frac{3}{4}\mu_\eta
+\frac{2B_0}{f_0^2}\left(4(m_{\rm ud}+m_{\rm s}) L_5+8(2 m_{\rm ud}+ m_{\rm s})L_4 
\right)\right\}, 
\label{eq:chpt_fk}
\een
where we have six unknown LECs $B_0,f_0,L_{4,5,6,8}$.
$\mu_{\rm PS}$ denotes the chiral logarithm defined by
\ben
\mu_{\rm PS}=\frac{1}{16\pi^2}\frac{{\tilde m}_{\rm PS}^2}{f_0^2}
\ln\left(\frac{{\tilde m}_{\rm PS}^2}{\mu^2}\right),
\label{eq:chlog}
\een
where
\ben
{\tilde m}_\pi^2 = 2 {m_{\rm ud}}B_0,\;\;\;&
{\tilde m}_K^2 = ({m_{\rm ud}} +m_{\rm s})B_0,\;\;\; &
{\tilde m}_\eta^2 = \frac{2}{3}({m_{\rm ud}} +2 m_{\rm s})B_0
\een
with $\mu$ the renormalization scale.
The pseudoscalar meson decay constants are calculated 
with the nonperturbatively
$O(a)$-improved axial vector current\cite{ca}, though the renormalization
factor is perturbatively evaluated 
up to one-loop level\cite{z_pt,z_imp_pt}. 
We determine the LECs  
by a simultaneous fit of $m_\pi^2/(2 m_{\rm ud})$,
$m_K^2/(m_{\rm ud} +m_{\rm s})$, $f_\pi$ and $f_K$
including the finite size corrections at one-loop
level\cite{colangelo05}.

\begin{figure}[h!]
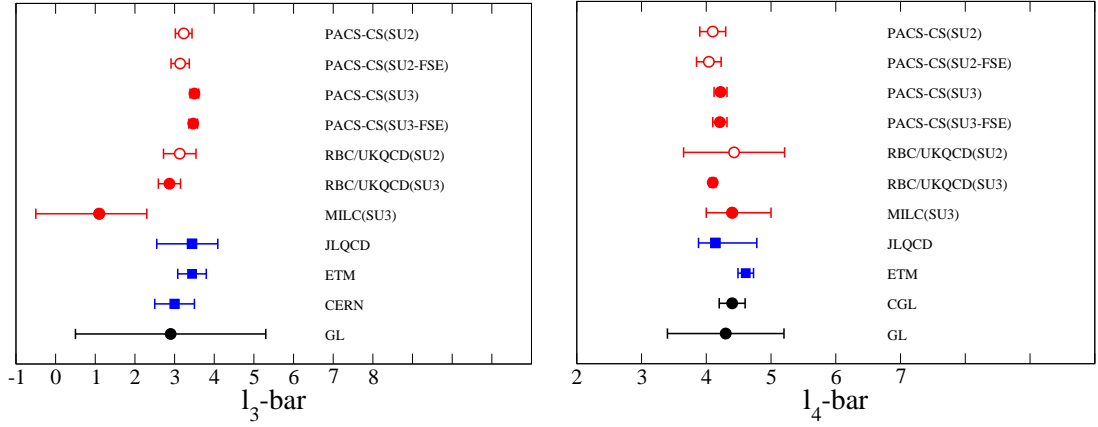

\vspace{3mm}
\begin{center}
\includegraphics[width=70mm,keepaspectratio,clip]{figs/lec_2f/l3-bar.eps}
\hspace{3mm}
\includegraphics[width=70mm,keepaspectratio,clip]{figs/lec_2f/l4-bar.eps}
\caption{Comparison of the results for $\bar l_3$ and $\bar l_4$. Black symbols
denote the phenomenological
estimates\cite{chpt_nf2,colangelo01}. Blue ones are for 2 flavor
 results\cite{del07,urbach07,jlqcd08}. Red closed (open) symbols represent the results for the SU(3) (SU(2)) ChPT
analyses in 2+1 flavor dynamical simulations\cite{milc07,rbcukqcd08}.  }
\label{fig:lec_2f}
\end{center}
\end{figure}

In Table~\ref{tab:lec_su3} we compare our results for the LECs with the
phenomenological estimates with experimental 
inputs\cite{colangelo05,amoros01} and recent
2+1 flavor lattice QCD results\cite{rbcukqcd08,milc07}.
We observe that the situation is rather complex: Some LECs are consistent
and others are not.
To make the comparison easier we convert the SU(3) LECs to the
SU(2) ones, where the number of LECs are reduced 
from six to four: $B$, $f$, ${\bar l}_3$, ${\bar l}_4$.
We obtain 
${\bar l}_3=3.47(11)$, ${\bar l}_4=4.21(11)$ and 
${\bar l}_3=3.50(11)$, ${\bar l}_4=4.22(10)$ with and without the finite
size corrections, respectively. 
These results are plotted in Fig.~\ref{fig:lec_2f} 
together with the phenomenological
estimates and the recent 2 and 2+1 flavor lattice QCD results.
We find that all the results reside in 
$3.0\simlt {\bar l}_3\simlt 3.5$ and $4.0\simlt {\bar l}_4\simlt 4.5$ 
except the MILC result for ${\bar l}_3$ which shows exceptionally small value.

\begin{table*}[h!]
\centering
\caption{Results for the LECs in the SU(3) ChPT 
together with the
phenomenological estimates\cite{colangelo05,amoros01} 
and the RBC/UKQCD\cite{rbcukqcd08} and the MILC results\cite{milc07}.
$f_0$ is perturbatively renormalized at one-loop level.
$L_{4,5,6,8}$ are in units of $10^{-3}$ at the scale of 770MeV.}
\label{tab:lec_su3}
\begin{tabular}{cccccc}
\hline
& \multicolumn{2}{c}{PACS-CS} & phenomenology\cite{colangelo05,amoros01} & 
 RBC/UKQCD\cite{rbcukqcd08} & MILC\cite{milc07} \\
& w/o FSE  & w/ FSE   &  & &   \\
\hline
$f_0$[GeV]  & 0.1160(88) & 0.1185(90) & 0.115 & 0.0935(73) & $-$ \\
$f_\pi/f_0$  & 1.159(57) & 1.145(56) & 1.139 &  1.33(7)   &  $1.21(5)\left(^{+13}_{-3}\right)$ \\
$L_4$        & $-0.04(10)$  & $-0.06(10)$       & 0.00(80)  
& 0.139(80) & $0.1(3)\left(^{+3}_{-1}\right)$           \\
$L_5$        & 1.43(7)  & 1.45(7)       & 1.46(10)  
& 0.872(99) & $1.4(2)\left(^{+2}_{-1}\right)$           \\
$2L_6-L_4$   &  0.10(2)  &  0.10(2)         & 0.0(1.0)     
& $-$0.001(42) & $0.3(1)\left(^{+2}_{-3}\right)$ \\ 
$2L_8-L_5$    &  $-0.21(3)$  & $-0.21(3)$      & 0.54(43)  
& 0.243(45) & 0.3(1)(1)    \\
\hline
$\chi^2$/dof & 4.2(2.7) & 4.4(2.8)  & $-$ & 0.7 & $-$  \\
\hline
\end{tabular}
\end{table*}

\begin{figure}[h!]
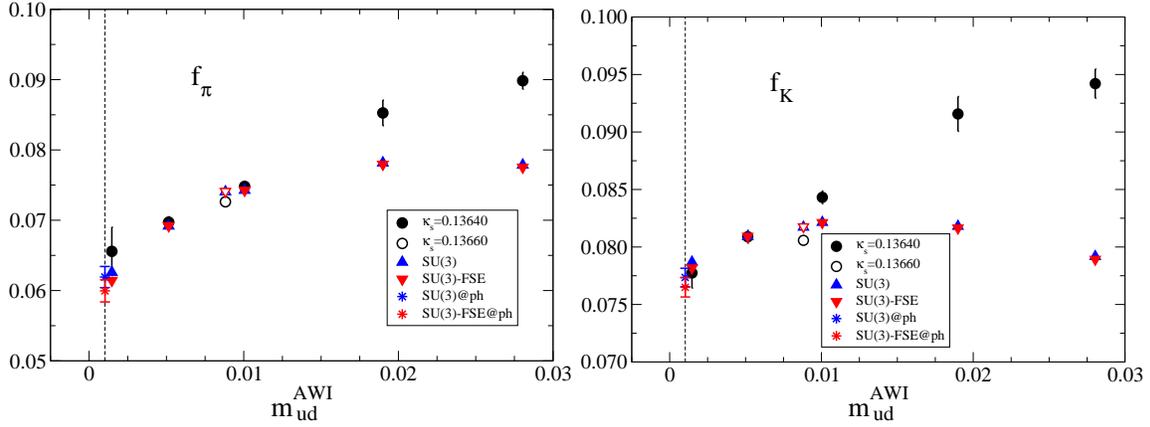

\begin{center}
\includegraphics[width=75mm,keepaspectratio,clip]{figs/chfit_su3/fpi.eps}
\includegraphics[width=75mm,keepaspectratio,clip]{figs/chfit_su3/fk.eps}
\caption{Fit results for $f_\pi$
 (left) and $f_K$ (right). Black  symbols 
represent the lattice results. 
Red triangles denote the SU(3) ChPT results plotted
at the measured quark masses. 
Open and filled symbols distinguish between the results 
at $\kappa_{\rm s}=0.13660$ and 0.13640. 
Star symbols represent the extrapolated values at the physical point
which is denoted by vertical dotted line. }
\label{fig:chfit_su3}
\end{center}
\end{figure}

\begin{figure}[h!]
\begin{center}
\includegraphics[width=80mm,keepaspectratio,clip]{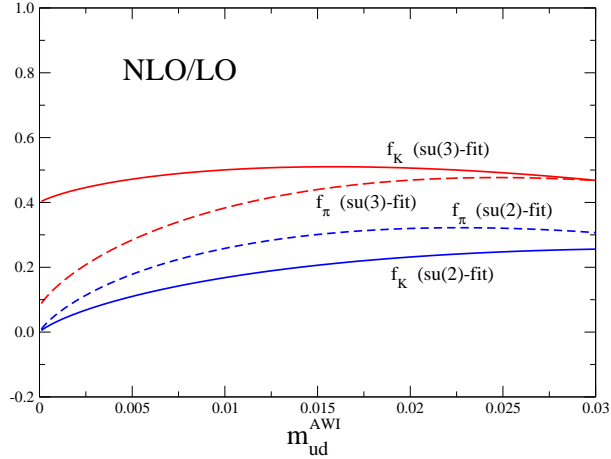}
\caption{Ratio of the NLO contribution to the LO one in the chiral
 expansions of the pseudoscalar meson decay constants.
Solid (open) lines represent the SU(3) (SU(2)) ChPT case.
The strange quark mass is fixed at the physical value. }
\label{fig:chpt_nlo2lo}
\end{center}
\end{figure}

Although we obtain reasonable values for ${\bar l}_3$ and ${\bar l}_4$
in the SU(3) ChPT fit, there exists a concern that 
the values of $\chi^2$/dof are unacceptably large. 
Its origin is found in the fit of $f_\pi$ and $f_K$ depicted 
in Fig.~\ref{fig:chfit_su3}, where
the strange quark mass dependence between the data at 
$(\kappa_{\rm ud},\kappa_{\rm s})=(0.13754,0.13640)$ and $(0.13754,0.13660)$
is not properly described by the SU(3) ChPT.
The convergence of the SU(3) ChPT is checked in 
Fig.~\ref{fig:chpt_nlo2lo}.
For $f_K$ the NLO contribution is about 40\% of the LO one 
almost independent of the up-down quark mass.  
This implies that the strange quark mass is not small enough to be
treated in the SU(3) ChPT up to NLO.
However, the extension to NNLO is not a good choice in a
practical sense:
The NNLO contribution could be 20\% by naive power counting
and a full determination of additional LECs is not achieved 
without increasing the data points significantly.
We take a different direction: the SU(2) ChPT fit 
with the analytical expansion
in terms of the strange quark mass about the physical value.

\subsection{SU(2) ChPT}
\label{subsec:chpt_su2}

We try further chiral analyses on the pseudoscalar meson sector 
employing the SU(2) ChPT up to NLO, where the $K$ meson is treated
as a matter field in the isospin 1/2 linear representation of the SU(2)
chiral transformation\cite{roessl}.  
The strange quark
contributions to the SU(2) LECs are analytically 
expanded about the physical value.
The SU(2) ChPT formulae for $m_\pi$ and $f_\pi$ are given by 
\ben
\frac{m_\pi^2}{2{m_{\rm ud}}}&=&B\left\{1+
\frac{1}{16\pi^2}\frac{{\bar m}_\pi^2}{f^2}
\ln\left(\frac{{\bar m}_\pi^2}{\mu^2}\right)+
4\frac{{\bar m}_\pi^2}{f^2}l_3\right\},
\label{eq:su2chpt_mpi}\\
f_\pi&=&f\left\{1- \frac{1}{8\pi^2}\frac{{\bar m}_\pi^2}{f^2}
\ln\left(\frac{{\bar m}_\pi^2}{\mu^2}\right)
+2\frac{{\bar m}_\pi^2}{f^2}l_4\right\}
\label{eq:su2chpt_fpi}
\een
with ${\bar m}_\pi^2=m_{\rm ud}B$.
$B$ and $f$ are linearly expanded in terms of the strange quark mass:
$B=B_s^{(0)}+m_{\rm s}B_s^{(1)}$ and 
$f=f_s^{(0)}+ m_{\rm s}  f_s^{(1)}$.
For $m_K$ and $f_K$ we employ the following fit formulae:
\ben
m_K^2=\bar m_K^2+\beta_m m_{\rm ud}, \qquad
f_K=\bar f \left\{1+\beta_f m_{\rm ud} 
- \frac{3}{4} \frac{2 B m_{\rm ud}}{16\pi^2 f} {\rm ln} \left(\frac{2B m_{\rm ud}}{\mu^2} \right)\right\},
\een
where $\bar m_K^2$ and $\bar f$ are also linearly expanded 
in terms of the strange quark mass: $\bar m_K^2=\alpha_m+\gamma_m m_{\rm
s}$ and 
$\bar f=\bar f_s^{(0)}+ m_{\rm s} \bar f_s^{(1)}$.

\begin{figure}[h!]
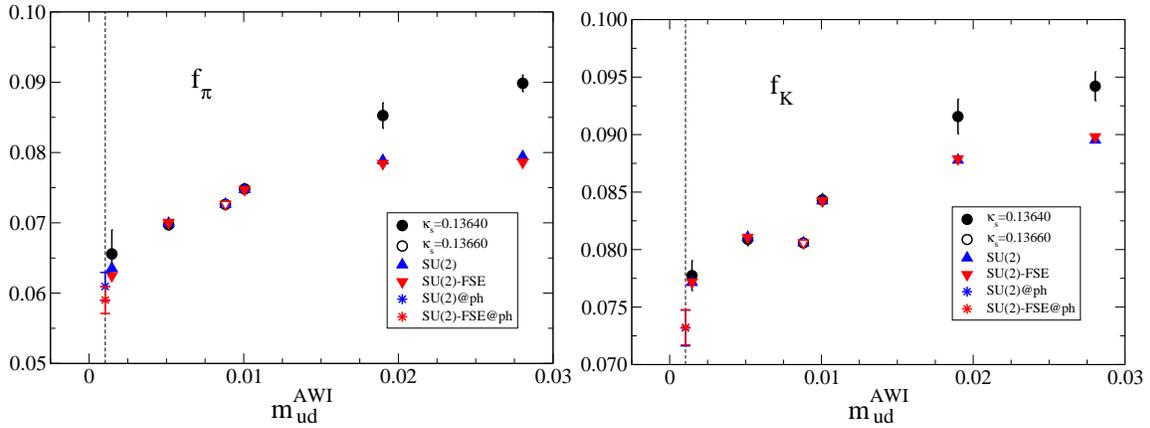

\begin{center}
\includegraphics[width=75mm,keepaspectratio,clip]{figs/chfit_su2/fpi.eps}
\includegraphics[width=75mm,keepaspectratio,clip]{figs/chfit_su2/fk.eps}
\caption{Same as Fig.~2 for the SU(2) ChPT.}
\label{fig:chfit_su2}
\end{center}
\end{figure}

A simultaneous fit to $m_\pi^2/2m_{\rm ud}$, $f_\pi$ and $f_K$ leads to
the results of 
${\bar l}_3=3.14(23)$, ${\bar l}_4=4.04(19)$ and 
${\bar l}_3=3.23(21)$, ${\bar l}_4=4.10(20)$ with and without the finite
size corrections, respectively. We apply an independent fit to 
$m_K^2$. The results for ${\bar l}_3$ and ${\bar l}_4$ 
 are added in 
Fig.~\ref{fig:lec_2f},  where we find  that all the lattice results 
are consistent except the MILC result of ${\bar l}_3$.
The fit results for $f_\pi$ and $f_K$ are depicted in Fig.~\ref{fig:chfit_su2}.
The lattice data are well described by the SU(2) ChPT
combined with the analytic expansion in terms of the strange quark mass.
The corresponding $\chi^2$/dof are 0.43(77) and 0.33(68) with and
without the finite size corrections, respectively, which are reasonable
magnitude contrary to the SU(3) case. 
In Fig.~\ref{fig:chpt_nlo2lo} we compare the convergence behaviors
between the SU(3) and the SU(2) cases. The ratio of the NLO contribution
to the LO one in the SU(2) ChPT is drastically reduced from the SU(3)
case in the range of $m_{\rm ud}\simlt 0.01$.

\section{ChPT analyses on the baryon sector}

It is intriguing to investigate the quark mass dependence of the
baryon mass based on the heavy baryon ChPT (HBChPT).
Let us consider the nucleon mass as an example.
The SU(2) heavy baryon ChPT 
formula up to NNLO is given by\cite{hbchpt}
\ben
&&m_N=m_0-4c_1 m_\pi^2 -\frac{6g_A^2}{32\pi^2f_\pi^2}m_\pi^3 \nonumber \\
&&\hspace{1cm} +\left[e_1(\mu)-\frac{6}{64\pi^2f_\pi^2}\left(\frac{g_A^2}{m_0}-\frac{c_2}{2}\right)
-\frac{6}{32\pi^2f_\pi^2}\left(\frac{g_A^2}{m_0}-8c_1+c_2+4c_3\right) {\rm ln}{\left(\frac{m_\pi}{\mu}\right)}
\right]m_\pi^4 \nonumber \\
&&\hspace{1cm} +\frac{6g_A^2}{256 \pi f_\pi^2 m_0^2}m_\pi^5 + O(m_\pi^6)
\label{eq:nucleon_chpt}
\een
with $m_0,f_\pi,c_1,c_2,c_3,g_A, e_1$ low energy constants.
Note that the inclusion of the $O(m_\pi^5)$ term, which is obtained by the
relativistic baryon ChPT\cite{hbchpt_rel}, little affects on the
following analyses because of its tiny contribution.

The HBChPT fit is performed under several constraints.
Firstly, the fit parameters are restricted to $m_0$, $c_1$, $e_1$.
For a direct comparison with the 2 flavor results of the QCDSF-UKQCD
Collaborations\cite{qcdsfukqcd_n}, 
we choose $g_A=1.267,c_2=3.2{\rm GeV}^{-1}$ with $c_3=-3.4{\rm
GeV}^{-1}$ or $c_3=-4.7{\rm GeV}^{-1}$. The former is refereed to as Fit-A
and the latter as Fit-B. We employ the value of $f_\pi$ at the chiral limit
obtained by the SU(2) ChPT fit for $m_\pi, f_\pi$ and $f_K$.
Secondly, the ChPT fit is applied to the data series 
at $\kappa_{\rm s}=0.13640$ which
show a few \% level of variation for $m_{\rm s}^{\rm AWI}$. We take two fit
ranges: $0.13781\le \kappa_{\rm ud} \le 0.13727$ for Range-I and
$0.13781\le \kappa_{\rm ud} \le 0.13700$ for Range-II.

\begin{figure}[h!]
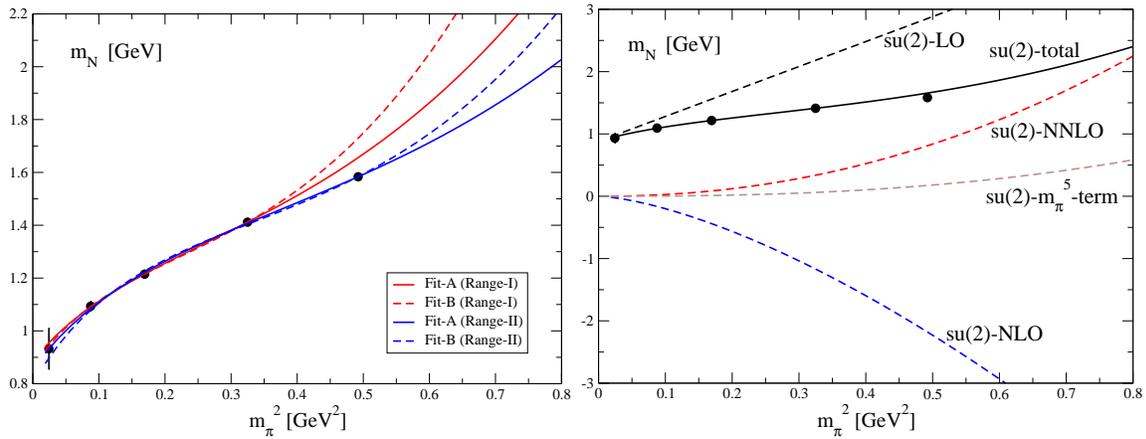

\begin{center}
\includegraphics[width=75mm,keepaspectratio,clip]{figs/nucleon_nf2/mpi2.mN.eps}
\includegraphics[width=75mm,keepaspectratio,clip]{figs/nucleon_nf2/ratio.mN.eps}
\caption{Fit results of the nucleon mass using the SU(2) heavy baryon
ChPT formula
(left) and each contribution of the
 LO, NLO, NNLO and $O(m_\pi^5)$ terms in the case of Fit-A with Range-I 
(right).}
\label{fig:chpt_n}
\end{center}
\end{figure}

Table~\ref{tab:chpt_n} presents the fit results for the LECs in comparison
with the QCDSF/UKQCD results\cite{qcdsfukqcd_n}. We also list 
the ETM results obtained
at $\beta=3.9$, which yields similar lattice spacing to ours, 
with the choice of $c_3=-3.45{\rm GeV}^{-1}$\cite{etm_n}. 
The results show reasonable agreement
in case that we take the similar values for $c_3$, 
We also give 
the nucleon sigma term evaluated through 
$\sigma_{N\pi}=m_\pi^2(\partial m_N/\partial m_\pi^2)$ with the aid of
 Eq.(\ref{eq:nucleon_chpt}).  
In Fig.~\ref{fig:chpt_n} we draw the fit results and investigate each
contribution of the LO, NLO, NNLO and $O(m_\pi^2)$ terms 
in Eq.(\ref{eq:nucleon_chpt}).
It is remarkable that the lattice results are
well described up to $m_\pi^2\sim 0.5$ GeV$^2$ both for Fit-A and Fit-B.
We also find reasonable values for $\chi^2$/dof 
in Table~\ref{tab:chpt_n}. 
However, this is achieved by a drastic cancellation between the LO and
NLO contributions. It is hard to say that the convergence of the SU(2) 
HBChPT is controlled beyond $m_\pi^2\sim 0.2$ GeV$^2$.

\begin{table}[h!]
\begin{center}
\caption{Fit results for  $m_0, c_1, e_1$ in comparison 
with the QCDSF/UKQCD and the  ETM results. 
Values for the nucleon sigma term are also presented.}  
\label{tab:chpt_n}
\begin{tabular}{cccccccc}\hline
           &  \multicolumn{2}{c} {Range-I} & \multicolumn{2}{c}
 {Range-II}  \\
                           & Fit-A & Fit-B & Fit-A & Fit-B   \\
$c_3$ [GeV$^{-1}$]  & $-3.4$ & $-4.7$ &  $-3.4$ & $-4.7$  \\
\hline
$m_0$ [GeV]                 & 0.880(50)   & 0.855(47)   & 0.850(27) & 0.795(27)  \\
$c_1$  [GeV$^{-1}$]             & $-$1.00(10)  & $-$1.19(10)   &
 $-$1.08(4) & $-$1.34(4) \\
$e_1(1{\rm GeV})$ [GeV$^{-3}$]      & 3.7(1.4)  & 4.2(1.4)   &
 2.9(4)  & 2.4(3)  \\   
\hline
$\chi^2$/dof         & 0.1(0.9) & 0.0(0.5)  & 0.3(0.9) & 1.2(1.6)  \\ 
\hline
$\sigma_{N\pi} $ [MeV]  & 51.4(7.6)   & 60.1( 7.3)   &  56.8(2.7) & 70.8(2.6) \\
\hline
\end{tabular}
\begin{tabular}{cccc}
&&& \\
\hline
           & \multicolumn{2}{c} {QCDSF-UKQCD\cite{qcdsfukqcd_n}} & ETM\cite{etm_n}\\
           & Fit-A & Fit-B & \\
$c_3$ [GeV$^{-1}$]  & $-3.4$ & $-4.7$ & $-3.45$ \\
\hline
$m_0$ [GeV]                    & 0.89(6)   & 0.76(6)  & 0.887(67)   \\
$c_1$  [GeV$^{-1}$]            & $-$0.93(10)  & $-$1.25(10) &  $-1.22(20)$   \\
$e_1(1{\rm GeV})$ [GeV$^{-3}$] &  2.8(1.4)  & 1.7(0.5) & $-$          \\   
\hline
\end{tabular}
\end{center}
\end{table}

\section{Results at the physical point}

\begin{figure}[t]
\vspace{3mm}
\begin{center}
\begin{tabular}{c}
\includegraphics[width=80mm,angle=0]{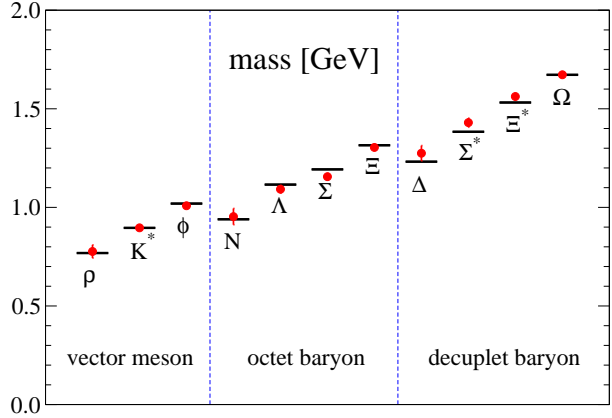}
\end{tabular}
\end{center}
\caption{Light hadron spectrum extrapolated to the physical point (red
 circles) in comparison with the experimental values (black bars).}
\label{fig:spectrum}
\end{figure}

We choose $m_\pi$, $m_K$ and $m_\Omega$ as physical inputs 
to determine the up-down and the
strange quark masses and the lattice cut-off.
The SU(2) ChPT analyses are applied to the quark mass dependences of
$m_\pi,f_\pi$ and $f_K$ taking account of the finite size
corrections evaluated at the one-loop level. 
We assume a simple linear quark mass dependences for $m_K^2$.
These points are already explained in Sec.~\ref{subsec:chpt_su2}.
We also employ linear forms for the chiral fits of the vector
and the baryon masses. It is difficult to rely on the heavy meson
effective theory\cite{hmet} and the heavy baryon ChPT because of their bad 
convergence behaviors in the chiral expansions\cite{convergence}.
Figure~\ref{fig:spectrum} compares
the light hadron spectrum at the physical point 
with the experimental values. The results are encouraging.
The largest discrepancy is at most 3\%, which may be explained
by possible $O((a\Lambda_{\rm QCD})^2)$ cutoff errors.
What we are left with is a proper analysis of 
the $\rho$ meson and the $\Delta$ baryon as resonance states.

We calculate the bare quark masses using the axial vector Ward-Takahashi
identity (AWI) with the nonperturbatively
$O(a)$-improved axial vector current\cite{ca}. The renormalization
factors $Z_A$ and $Z_P$ are perturbatively evaluated 
up to one-loop level\cite{z_pt,z_imp_pt}. In Table~\ref{tab:qmass_phpt} we
present the results for the quark masses, the lattice cutoff and the
pseudoscalar meson decay constants together with the experimental values
and the recent 2+1 flavor lattice QCD results.
The physical up-down quark mass is just 30\% smaller than our lightest
one $m_{\rm ud}^{\msbar}(\mu=1/a)=3.5$ MeV at $(\kappa_{\rm
ud},\kappa_{\rm s})=(0.13781,0.13640)$. We find that the RBC/UKQCD 
results for the physical quark masses give rather larger values
than ours. This may be caused by different renormalization procedures:
a nonperturbative implementation of the RI-MOM scheme for the former 
and one-loop perturbation for the latter. In fact, the ratio of the
up-down quark mass to the strange one, which should be independent of
the renormalization factors, shows a good agreement. It may be also 
suggestive that 
the MILC results obtained with two-loop renormalization factors are
found in the middle of ours and the RBC/UKQCD ones. We are now
evaluating the nonperturbative renormalization factors employing
the Schr{\"o}dinger functional scheme.

In Table~~\ref{tab:qmass_phpt} we also present the results 
for the pseudoscalar meson decay constants
at the physical point employing the perturbative renormalization factor
at one-loop level. We find that our results are consistent 
with the experimental values within the errors. 
On the other hand, the RBC/UKQCD results 
are smaller than the experimental values by 5\%, albeit rather
large systematic uncertainties are estimated. 
Since the MILC results are free from
the uncertainties due to the finite renormalization, 
they use $f_\pi$ as a physical input.

\begin{table*}[h!]
\centering
\caption{Cutoff, renormalized 
quark masses, pseudoscalar meson decay constants determined with 
$m_\pi$, $m_K$ and $m_\Omega$ as physical inputs. 
Quark masses are renormalized at 2 GeV.}
\label{tab:qmass_phpt}
\begin{tabular}{cccccc}  
\hline
 &  \multicolumn{2}{c}{physical point} & experiment\protect{\cite{pdg}}
 &  RBC/UKQCD\cite{rbcukqcd08} & MILC\cite{milc07} \\ 
 &   w/o FSE &  w/ FSE  &  \\ 
\hline 

$a^{-1}$~[GeV]  & 2.176(31) & 2.176(31) & $-$  & 1.729(28) & {\rm continuum}\\ 
$m^{\overline{\rm MS}}_{\rm ud}$~[MeV] & 2.509(46) & 2.527(47) & $-$ & 3.72(16)(33)(18)&
 3.2(0)(1)(2)(0) \\ 
$m^{\overline{\rm MS}}_{\rm s}$~[MeV]  & 72.74(78) & 72.72(78) & $-$ &
 107.3(4.4)(9.7)(4.9) & 88(0)(3)(4)(0) \\ 
$m_{\rm s}/m_{\rm ud}$ & 29.0(4) & 28.8(4) & $-$  & 28.8(0.4)(1.6) & 27.2(1)(3)(0)(0)\\
$f_\pi$~[MeV] & 132.6(4.5) & 134.0(4.2) & $130.7\pm 0.1\pm 0.36$  &
 124.1(3.6)(6.9) & {\rm input}\\ 
$f_K$~[MeV]   & 159.2(3.2) & 159.4(3.1) & $159.8\pm 1.4\pm 0.44$  &
 149.6(3.6)(6.3) & 156.5(0.4)$(^{+1.0}_{-2.7})$ \\ 
$f_K/f_\pi$   & 1.201(22)  & 1.189(20)  & 1.223(12)  & 1.205(18)(62) & 1.197(3)$(^{+6}_{-13})$\\ 
\hline
\end{tabular}
\end{table*}

\section{Results of the physical point simulation}

A direct simulation at the physical point is attempted by
choosing the hopping parameters as $(\kappa_{\rm ud},\kappa_{\rm
s})=(0.137785,0.13660)$, which is estimated using the four data set
with $\kappa_{\rm ud}\ge 0.13754$.  
The measured up-down and the strange quark masses are 
\begin{eqnarray}
m_{\rm ud}^{\overline{\rm MS}}(\mu=2{\rm GeV})=3.59(20){\rm MeV},&&
m_{\rm s}^{\overline{\rm MS}}(\mu=2{\rm GeV})=73.72(44){\rm MeV},
\end{eqnarray}
where $a^{-1}$ is determined from $m_\Omega$ at  
$(\kappa_{\rm ud},\kappa_{\rm s})=(0.137785,0.13660)$.
The strange quark mass is successfully tuned at the
physical value given in Table~\ref{tab:qmass_phpt}.
Although the up-down quark mass is $5\sigma$ away from the physical value, 
it is shown that the chiral corrections due to the
deviation are negligibly small or comparable 
to the current statistical errors.
We discuss the importance of the physical point simulation 
from a view point of controlling the systematic errors.

\begin{figure}[t]
\vspace{3mm}
\begin{center}
\begin{tabular}{c}
\includegraphics[width=80mm,angle=0]{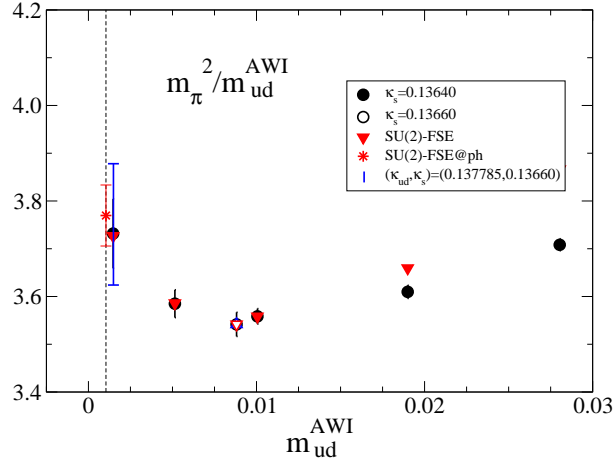}
\end{tabular}
\end{center}
\vspace{-.5cm}
\caption{Chiral extrapolation of 
$m_\pi^2/m_{\rm ud}^{\rm AWI}$ together with the result at
$(\kappa_{\rm ud},\kappa_{\rm s})=(0.137785,0.13660)$. SU(2) ChPT 
fit is applied to
four data points with $m_{\rm ud}^{\rm AWI}\simlt 0.01$.}
\label{fig:chexp_mpi}
\end{figure}

\begin{figure}[h!]
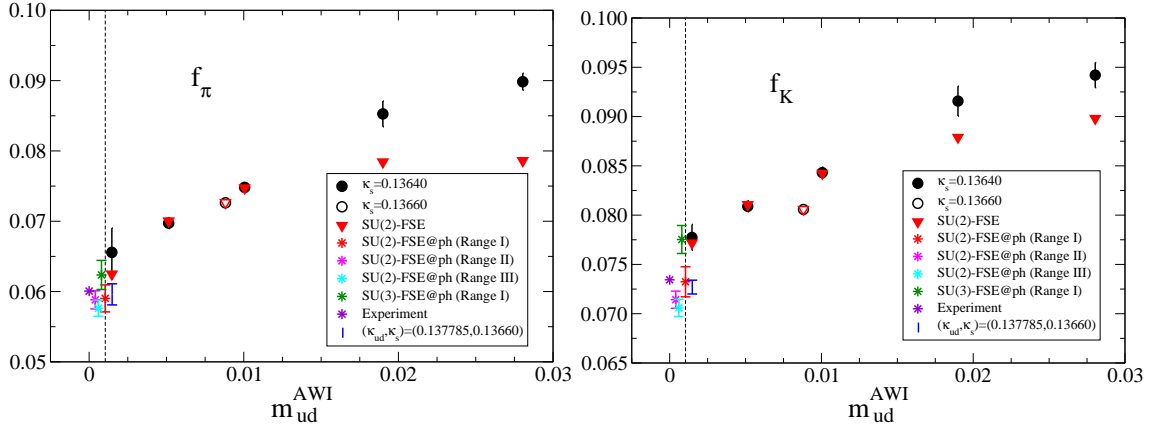

\begin{center}
\includegraphics[width=75mm,keepaspectratio,clip]{figs/phpt/fpi.eps}
\includegraphics[width=75mm,keepaspectratio,clip]{figs/phpt/fk.eps}
\caption{Chiral extrapolations of $f_\pi$ (left) and $f_K$ (right)
together with the results at
$(\kappa_{\rm ud},\kappa_{\rm s})=(0.137785,0.13660)$. Red symbols
represent the SU(2) ChPT 
fit applied to four data points with $m_{\rm ud}^{\rm AWI}\simlt 0.01$.}
\label{fig:chexp_fps}
\end{center}
\end{figure}

We first investigate the pseudoscalar meson sector.
In Fig.~\ref{fig:chexp_mpi} we plot $m_\pi^2/m_{\rm ud}^{\rm AWI}$ 
as a function of $m_\pi^2/m_{\rm ud}^{\rm AWI}$  
together with the fit results 
with the SU(2) ChPT. The result at $(\kappa_{\rm ud},\kappa_{\rm
s})=(0.137785,0.13660)$ shows an agreement with 
the extrapolated value at the physical point (red star) 
within rather large statistical error. 
Figure~\ref{fig:chexp_fps} illustrates an importance of the physical point
simulation. For $f_\pi$ we observe that the extrapolated values at the
physical point with various fit formulae are consistent within the error
bars. The story is different for $f_K$:  
The extrapolated values are scattered beyond the error bars. 
This difference is caused by the typical
magnitude of the error bar: 4\% for $f_\pi$ 
and $1-2$\% for $f_K$. The increase of the resolution with the diminishing
statistical errors reveals the discrepancies between the various chiral
extrapolations. Figure~\ref{fig:chexp_fps} tells us that 
these uncertainties should be removed   
by a direct simulation at the physical point.
In the pseudoscalar meson sector it is clear that the up-down quark mass 
at $(\kappa_{\rm ud},\kappa_{\rm s})=(0.137785,0.13660)$
is so close to the physical value that the chiral corrections from the
physical point are not relevant within 
the current statistical errors.

\begin{figure}[t]
\vspace{3mm}
\begin{center}
\begin{tabular}{c}
\includegraphics[width=80mm,angle=0]{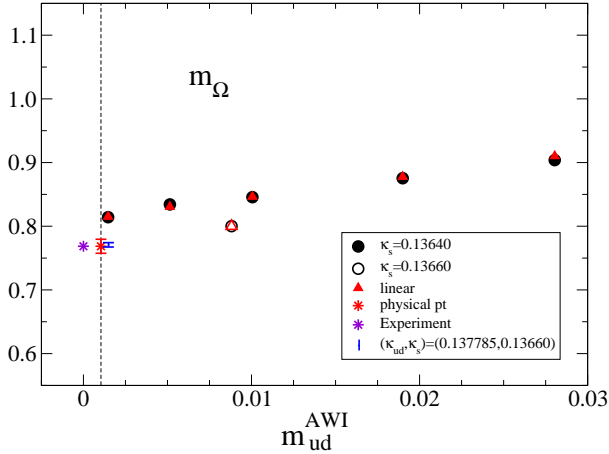}
\end{tabular}
\end{center}
\vspace{-.5cm}
\caption{Chiral extrapolation of $m_\Omega$
together with the result at
$(\kappa_{\rm ud},\kappa_{\rm s})=(0.137785,0.13660)$. Linear fit is applied to
four data points with $m_{\rm ud}^{\rm AWI}\simlt 0.01$.}
\label{fig:chexp_momega}
\end{figure}

Since we employ $m_\Omega$ as a physical input to determine the lattice
cutoff, it is intriguing to check the result of $m_\Omega$ at $(\kappa_{\rm
ud},\kappa_{\rm s})=(0.137785,0.13660)$.
In Fig.~\ref{fig:chexp_momega} we compare it with the extrapolated 
value at the physical point. We find a good agreement between them.
Moreover, the result at $(\kappa_{\rm
ud},\kappa_{\rm s})=(0.137785,0.13660)$ has a smaller error than 
the extrapolated value. In this case the chiral corrections from the
the physical point could be comparable to the statistical error.
It is fascinating that we are allowed to make a precise measurement of 
the physical input directly at the physical point.

\begin{figure}[h!]
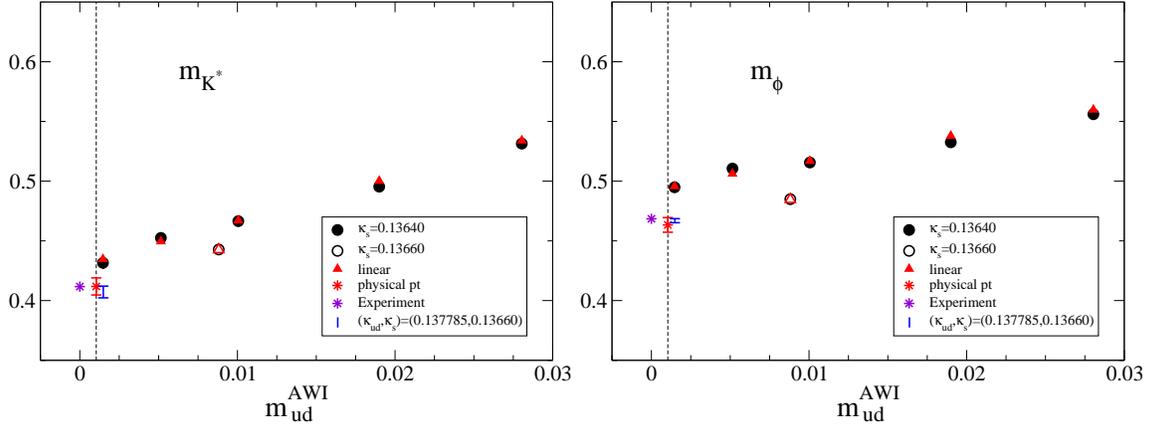

\begin{center}
\includegraphics[width=75mm,keepaspectratio,clip]{figs/phpt/V_LS.eps}
\includegraphics[width=75mm,keepaspectratio,clip]{figs/phpt/V_SS.eps}
\caption{Same as Fig.~9 for $m_{K^*}$ (left) and $m_\phi$ (right).}
\label{fig:chexp_mv}
\end{center}
\end{figure} 

For the vector channel we compare the results for $m_{K^*}$ and $m_\phi$
at  $(\kappa_{\rm ud},\kappa_{\rm s})=(0.137785,0.13660)$ with
the extrapolated values in Fig.~\ref{fig:chexp_mv}.
The situations are similar to the $m_\Omega$ case:
The extrapolated values are confirmed by the direct measurements at  
$(\kappa_{\rm ud},\kappa_{\rm s})=(0.137785,0.13660)$ with the smaller
statistical errors.

\section*{Acknowledgment}
Numerical calculations for the present work have been carried out
on the PACS-CS computer
under the ``Interdisciplinary Computational Science Program'' of
Center for Computational Sciences, University of Tsukuba.
This work is supported in part by Grants-in-Aid for Scientific
Research from the Ministry of Education, Culture, Sports, Science and
Technology (Nos.
16740147,   
17340066,   
18104005,   
18540250,   
18740130,   
19740134,   
20340047,   
20540248,   
20740123,   
20740139    
).

\end{document}